\documentclass[a4paper,12pt]{article}\pdfoutput=1
\usepackage[english]{babel}
\usepackage{amsmath}
\usepackage{amsfonts}
\usepackage{amssymb}
\usepackage{bbm}
\usepackage{graphicx}
\usepackage{hyperref}

\graphicspath{{pics/}}

\newcommand{\set}[1]{\left\{#1\right\}}
\newcommand{\hh}[1]{\left(#1\right)}
\newcommand{\sh}[1]{(#1)}
\newcommand{\abs}[1]{|#1|}
\newcommand{\norm}[1]{|#1|}
\newcommand{\map}[3]{#1: #2\rightarrow #3}
\newcommand{\RR}{\mathbbm{R}}
\newcommand{\CC}{\mathbbm{C}}
\newcommand{\ii}{\mathbbm{i}}
\newcommand{\ee}{\mathbbm{e}}
\newcommand{\Id}{\mathbbm{I}}
\newcommand{\ISO}[1]{\textrm{ISO}(#1)}

\DeclareMathOperator{\arctanh}{arctanh}
\DeclareMathOperator{\tr}{Tr}
\DeclareMathOperator{\re}{Re}
\DeclareMathOperator{\im}{Im}

\title{\normalsize \hfill ITP-UU-10/08  \\ \hfill SPIN-10/08
\\ \vskip 10mm \Large\bf Collisions in piecewise flat gravity in 3+1 dimensions}
\author{Maarten van~de~Meent}
\date{\normalsize Institute for Theoretical Physics and Spinoza Institute\\ Utrecht University\\
 P.O. Box 80.195, 3508 TD Utrecht, the Netherlands\\
 \smallskip e-mail: \tt M.vandeMeent@uu.nl}

\hypersetup{pdfauthor={Maarten van de Meent},pdftitle={Collisions in piecewise flat gravity in 3+1 dimensions},pdfkeywords={ Locally flat - Straight strings - Holonomy - Classical general relativity - Exact solutions - Collisions}}
\begin{document}
\maketitle
\begin{abstract}
We consider the $(3+1)$-dimensional locally finite gravity model proposed by 't Hooft \cite{hooft2008}. In particular we revisit the problem of resolving collisions of string defects. We provide a new geometric description of the configurations of strings using piecewise flat manifolds, and use it to resolve a more general class of collisions. We argue that beyond certain bounds for the deficiency/surplus angles no resolutions may be found that satisfy the imposed causality conditions.
\end{abstract}

\section{Introduction}
In \cite{hooft2008} 't Hooft introduced a locally flat model for gravity in $3+1$ dimensions. The basic premise was an attempt to generalize some features of general relativity in $2+1$ dimensions to $3+1$ dimensions. In $2+1$ dimensions Einstein's equation prescribes that in the absence of matter spacetime is locally flat. Point particles can be introduced in $2+1$ dimensions as $(0+1)$-dimensional line defects producing a conical singularity along the line. Models like this have been studied at some length in the 1980's and 1990's. \cite{Deser:1983tn,'tHooft:1993gz,'tHooft:1996uc,Kadar:2004im,Witten:1988hc}

The position and orientation of the $(0+1)$-dimensional line and the magnitude of the singularity it produces are completely captured by the Poincar\'e holonomy it produces on a path looping the line defect. Alternatively, these may be thought of as the position, velocity, and energy of a point particle moving linearly through $(2+1)$-dimensional space. Each particle thus requires only a finite number of degrees of freedom to describe it, and since any finite volume of space contains only a finite number of particles this model locally has a finite number of parameters. This makes the model of interest for inquiries into quantum gravity.

The model introduced by 't Hooft revolves around lifting the properties of this $(2+1)$-dimensional model to $3+1$ dimensions. Normally, in $3+1$ dimensions Einstein's equation does not fix all curvature degrees of freedom in empty space. This freedom allows the propagation of gravitational waves and long distance gravitational fields. To reproduce the properties of the $(2+1)$-dimensional gravity we impose as an additional constraint on the curvature that space be flat in the absence of matter.\footnote{Note that, therefore, regions of spacetime that are Ricci flat but not Riemann flat, must be represented by dense configurations of strings as well.} The local degrees of freedom now turn out to be $(1+1)$-dimensional surfaces in spacetime, which may be interpreted as 1-dimensional straight strings moving through space at a constant velocity.

In sections \ref{sec:3+1d} we recall the $(3+1)$-dimensional string model introduced in \cite{hooft2008}. Section \ref{sec:collisions} discusses the new problem of collisions in that model and some of the basic results established by 't Hooft. In section \ref{sec:squareconfig} we review the quadrangle resolutions of such collisions suggested in \cite{hooft2008} and find a complete analytic solution. This solution confirms 't Hooft's conclusions based on numerical calculations that such a resolution cannot be made consistent with causality for extremely violent collisions. In section \ref{sec:tetrahedron} we introduce the more complicated tetrahedral resolutions and show that these can always be solved in the non-relativistic low energy limit.  In the general limit these resolutions become too complicated to solve with the previously employed algebraic methods. To deal with this we introduce an alternative geometric way of describing the string configurations based on piecewise flat manifolds in section \ref{sec:cellcomplexes}. Section \ref{sec:cellresolution} explains how to use this new method to obtain a tetrahedral resolution of a collision, while section \ref{sec:limits} discusses when these methods break-down and establishes that for the most violent collisions no resolution (of any type) that is consistent with causality can be found . Finally, section \ref{sec:comparison} discusses some of the similarities and differences with other gravity models that are piecewise flat.

\section{\texorpdfstring{Straight strings in $3+1$ dimensions}{Straight strings in 3+1 dimensions}}\label{sec:3+1d}
Imposing the conditions that Einstein's equation must apply and that empty space must be flat, implies that the only matter that can be introduced must appear as straight strings of constant density moving through space at a constant velocity. This follows from the fact that any curvature of the string or its path through space or any variation in its density would imply a non-zero curvature in the surrounding space through Einstein's equation.

The effect of a single such string standing still is easy enough to derive; it simply adds or removes a wedge of space parallel to the string. This effect can be recorded through the holonomy of a loop around the $(1+1)$-dimensional surface defined by the string. If the density of the string is positive the holonomy shows a deficit angle and if its density is negative it show a surplus angle.\footnote{The exact relation between the deficit angle $\alpha$ and the string mass density $\rho$ is given by $\alpha = \frac{8\pi G}{c^2}\rho$. We will employ units where $8\pi G=1$ and $c=1$ such that deficit angle and density become interchangeable.}

The effect of any other moving string can simply be obtained by taking a suitable Poincar\'e transformation of a stationary string. The (Poincar\'e) holonomy around the string will then contain all the information about the string. If $P$ is the holonomy of a string then the $(1+1)$-dimensional surface traced out by the moving string is given by $\set{x\in\RR^{3,1}|P(x)=x}$. The density of the string can be obtained (up to a sign) from the eigenvalues of $P$. 

Note that not all holonomies will describe a moving string. To see this consider an holonomy $P$, an element of the Poincar\'e group. By performing a coordinate shift, we can always bring it in the form of a pure Lorentz transformation $R\in SO(3,1)$. In order for the set $S=\set{x\in\RR^{3,1}|P(x)=x}$ ($=\set{x\in\RR^{3,1}|R(x)=x}$ in the shifted coordinates) to be two dimensional $R$ must have two eigenvalues equal to one. For the surface to be timelike as well, the eigenspace corresponding to eigenvalue one must contain a timelike vector. Thus when $S$ is a timelike surface, $R$ is a pure rotation in the frame in which it is block diagonal. It is easy to see the converse that when $R$ is a pure rotation in some frame, then $S$ is a $(1+1)$-dimensional (i.e. timelike) surface.   

Clearly not all Lorentz transformations $R$ have a frame in which they are a pure rotation.  In general, it can occur that $R$ has no timelike eigenvector.  In such a case the set $S=\set{x\in\RR^{3,1}|R(x)=x}$ will be spacelike. Such objects cannot  represent a physically propagating degree of freedom and should not appear in our model. 
 
\begin{figure}\centering
\includegraphics[width=80mm]{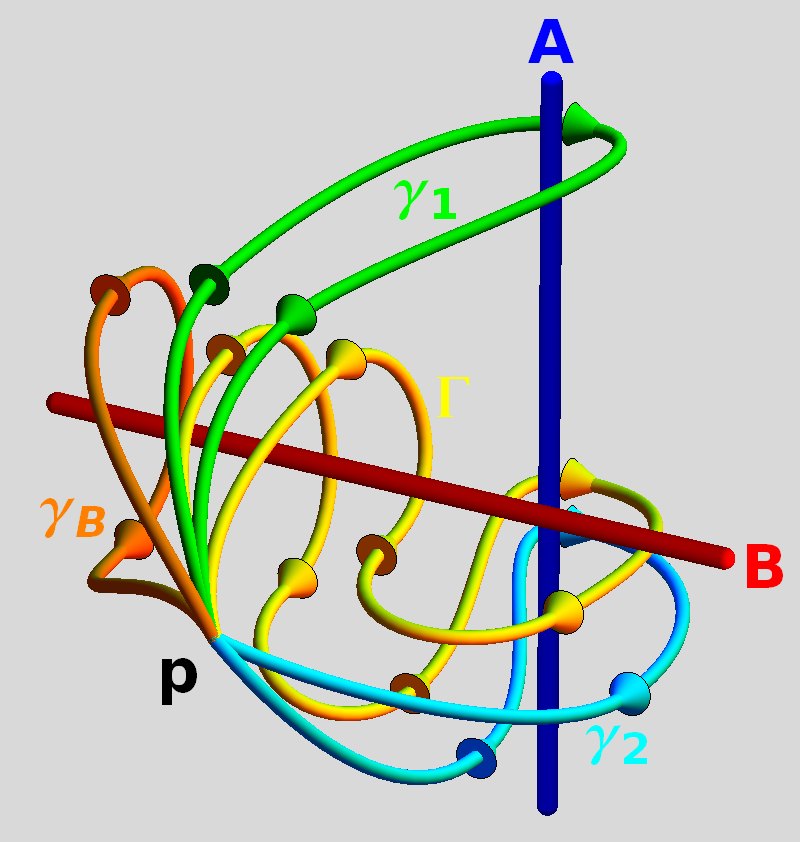}
\caption{The path $\gamma_1$ and path $\gamma_2$ around the string $A$ are topologically inequivalent. The path $\Gamma$ shows that $\gamma_1$ can be deformed to the sequence of paths $\gamma_B$ then $\gamma_2$ then $\gamma_B^{-1}$.}\label{fig:holoframes}
\end{figure}
More complex configurations of strings can similarly be described by specifying the holonomies around the various strings. Some care is needed however, since the holonomy of a string depends on the route taken to the string. For example consider the situation in figure \ref{fig:holoframes}. When describing a loop from point $p$ around string $A$ we have two topologically distinct options; we can either pass above ($\gamma_1$) or below ($\gamma_2$) string $B$.\footnote{Actually, there is an infinite number of options if we allow the paths to wrap around string $B$ before going to string $A$, but here we restrict our attention to these two options. The other options will lead to similar relations.} Topologically, the path $\gamma_1$ is equivalent to first going around string $B$ along path $\gamma_B$ then following path $\gamma_2$ around string $A$ and finally tracing back along $\gamma_B$. The holonomy of this last path is the product of the holonomies of the individual paths. Since the space away from the strings is flat the holonomy is a topological invariant of the path. Consequently, the holonomy of $\gamma_2$ should be equal to the holonomy of the combination of paths. So, if we denote the holonomies of the different paths $Q_{\gamma_1}$, $Q_{\gamma_2}$, and $Q_{\gamma_B}$, they should satisfy
\begin{equation}
Q_{\gamma_1} = Q_{\gamma_B}^{-1}Q_{\gamma_2}Q_{\gamma_B}.
\end{equation}
We see that $Q_{\gamma_1}$ is related to $Q_{\gamma_2}$ by a conjugation. This is true in general.  The holonomies of different paths wrapping around a string once belong to the same conjugacy class.

If we call our spacetime $X$ and the subset of $(1+1)$-dimensional string defects $X^{(2)}\subset X$, then the classes of topologically equivalent loops starting from the point $p\in X\setminus X^{(2)}$ are parameterized by the fundamental group $\pi_1\left[X\setminus X^{(2)},p\right]$, the group structure being defined by the concatenation of loops. To each class of loops there must be assigned a holonomy so we have a map
\begin{equation}
 \map{Q}{\pi_1\left[X\setminus X^{(2)},p\right]}{\ISO{T_p X}},
\end{equation}
that assigns to each loop $\gamma$ its holonomy $Q_\gamma$, an element of the Poincar\'e group at the point $p$: $\ISO{T_p X}$. The consistency requirement of the example above generalizes to the requirement that the map $Q$ respects the group structure of the groups, i.e. it is a group homomorphism.

As was the case for a single string, not any such homomorphism $Q$ will describe a configuration of strings. The holonomies assigned to loops wrapping around a single string must have a frame in which they are pure rotations. Note that if this condition is met for one path wrapping a string, it will be automatically met for all paths wrapping that string since these only differ by a change of frame.

Note also that this condition is not met for all loops. Loops wrapping multiple strings will generally not have a frame in which they are a pure rotation.

Thus far we have only discussed infinitely extended strings. That is $(1+1)$-dimensional string sheets that extend infinitely far along their space and time directions. Although one can build configurations consisting of only such strings, in general one would also expect two strings to meet. Generically, any two moving strings will collide at some point in spacetime.\footnote{Note that this collision can actually lie in the past of the strings. For ease of argument we will assume that the collision lies to the future of the considered configuration.} We will discuss how to deal with such events later, but to do so we must first introduce strings ending in junctions.

\begin{figure}\centering
\includegraphics[width=80mm]{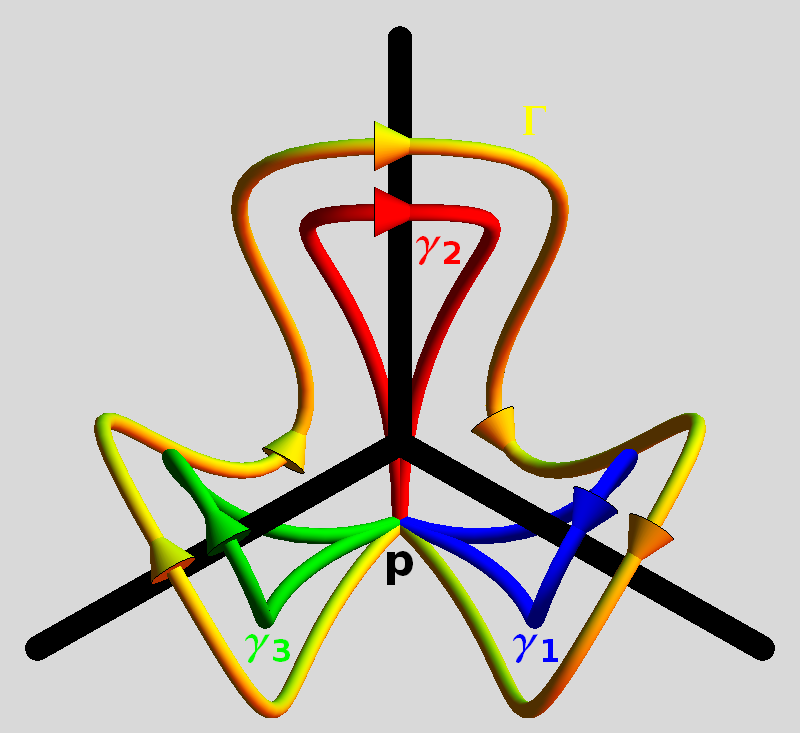}
\caption{The loop $\Gamma$ is homotopic to both the concatenation of paths $\gamma_1\cdot\gamma_2\cdot\gamma_3$ and to the trivial loop. The holonomies assigned to the loops $\gamma_i$ must thus satisfy $Q_{\gamma_1}Q_{\gamma_2}Q_{\gamma_3}=\Id$.}\label{fig:junction}
\end{figure}

A junction is a line in spacetime that is shared by multiple string sheets. That is it is a 1-dimensional set of points $x$ satisfying
\begin{equation}
Q_1 x=Q_2 x=...=Q_n x=x,
\end{equation}
with $Q_1$, ...,$Q_n$ being $n$ holonomies describing strings. Note that $n$ needs to be equal or larger than 3 in order for the surrounding spacetime to be flat.\footnote{That is a string cannot simply end somewhere in flat space, and if just two strings meet we are restricted to the trivial case that $Q_1 = Q_2^-1$.} Due to the presence of the junction the paths $\gamma_i$ defining the holonomies $Q_i$ will satisfy certain relations. For a simple example see figure \ref{fig:junction}, showing the equivalence of loops around three strings meeting in a junction. In general, one can always choose the paths $\gamma_i$ in such a way that the concatenation of paths $\gamma_1\cdot\dots\cdot\gamma_n$ is equivalent to the trivial loop. Since, the holonomies must follow the same algebraic conditions as the paths it follows that the holonomies must satisfy
\begin{equation}
Q_{\gamma_1}\cdots Q_{\gamma_n} = \Id.
\end{equation}
The 1-dimensional line of a junction can be either timelike, lightlike, or spacelike. Timelike and lightlike junctions simply represent  a point in space where multiple strings meet moving through space. Spacelike junctions, however, can be troublesome as they somehow represent information moving through spacetime at superluminal speeds. Such junctions are to be avoided by this model.

\section{Collisions}\label{sec:collisions}
When considering point particles in $2+1$ dimensions one could safely assume that the particles never collide, because generically the intersection of two $(0+1)$-dimensional lines in  $\RR^{2,1}$ is empty. When considering strings in $3+1$ dimensions, we have no such luxury since two $(1+1)$-dimensional surfaces in  $\RR^{3,1}$ generically have 1 intersection. That is any two strings moving though 3-dimensional space will collide at some point (unless they are parallel).

\begin{figure}
\centering
\includegraphics[width=120mm]{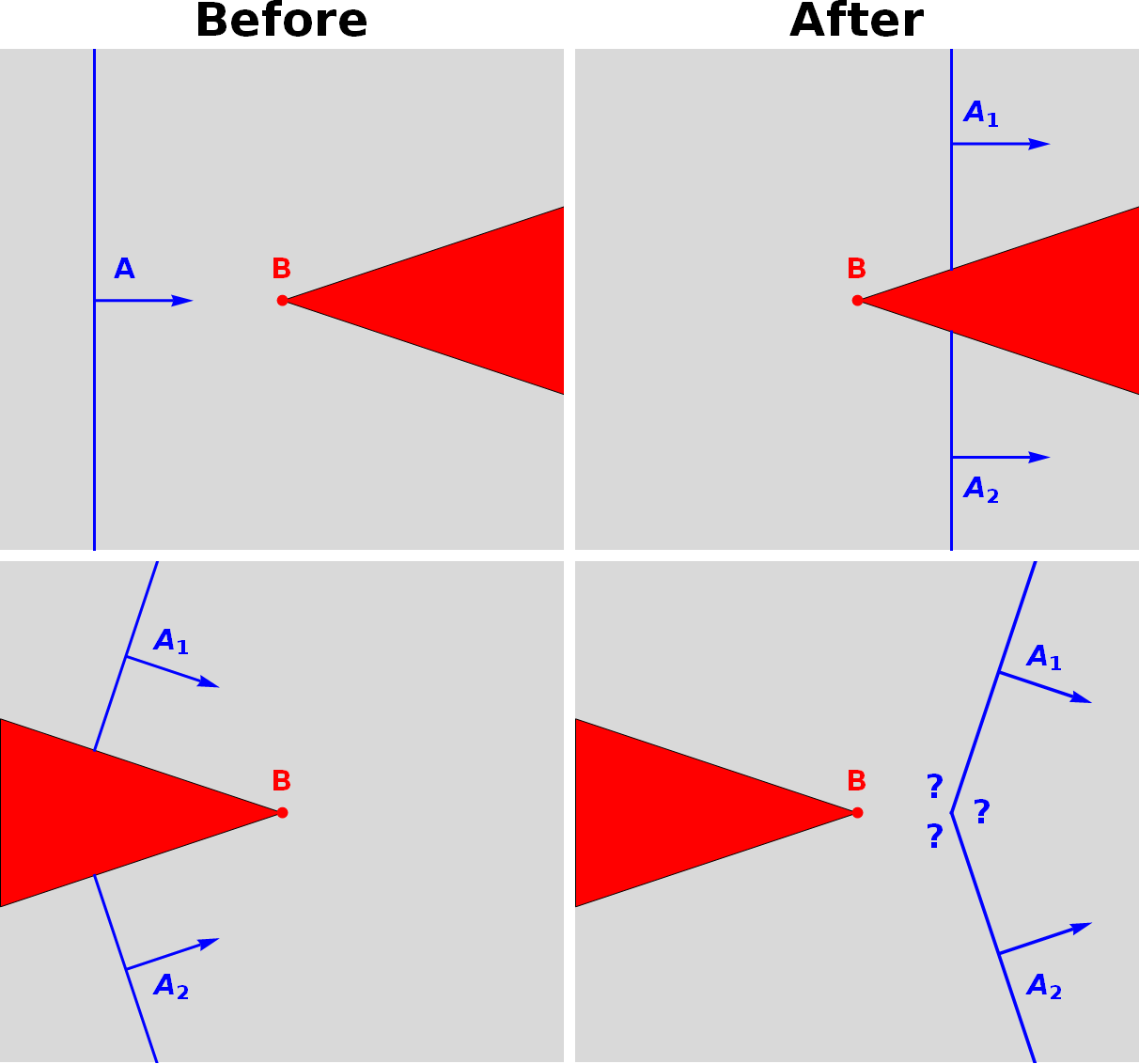}
\caption{String $A$ collides with string $B$ viewed along the direction of $B$. The red wedge represents the deficit angle produced by $B$.  The deficit angle can be removed in any direction from $B$. The top pictures show the situation before and after the collision of $A$ with $B$ with the deficit angle to the right of $B$. The bottom pictures show the same situation but with the deficit angle drawn to the left. We see that it is impossible for $A$ to continue as a straight line after colliding with $B$.}\label{fig:simplecollision}
\end{figure}

\begin{figure}
\centering
\includegraphics[width=80mm]{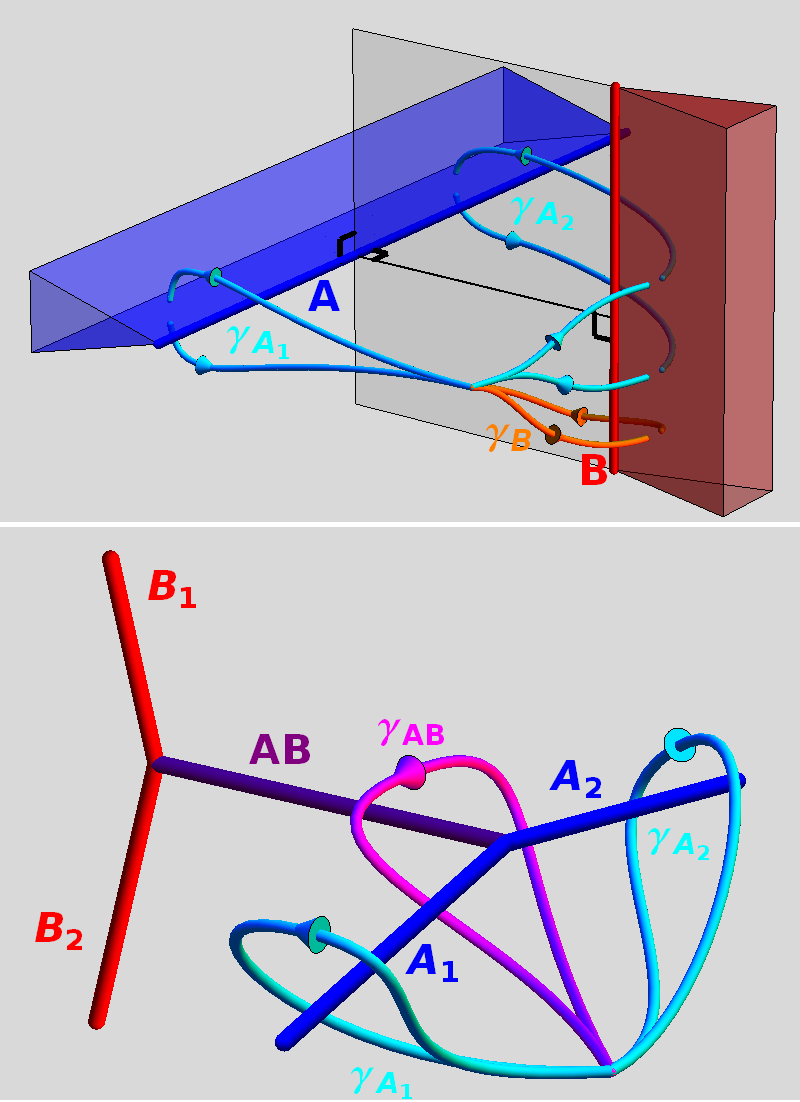}
\caption{Top: Two strings about to collide at a right angle. The loop $\gamma_{A_2}$ is equivalent to  $\gamma_{B}^{-1}\cdot\gamma_{A_1}\cdot\gamma_{B}$ Bottom: The same strings after the collision drawn without the wedges. The loops $\gamma_{A_1}$ and $\gamma_{A_2}$ are equivalent to the same loops in the top picture. The loop $\gamma_{AB}$ is equivalent to the loop $\gamma_{A_1}\cdot\gamma_{A_2}^{-1}.$}\label{fig:orthogonalcollision}
\end{figure}

\begin{figure}
\centering
\includegraphics[width=100mm]{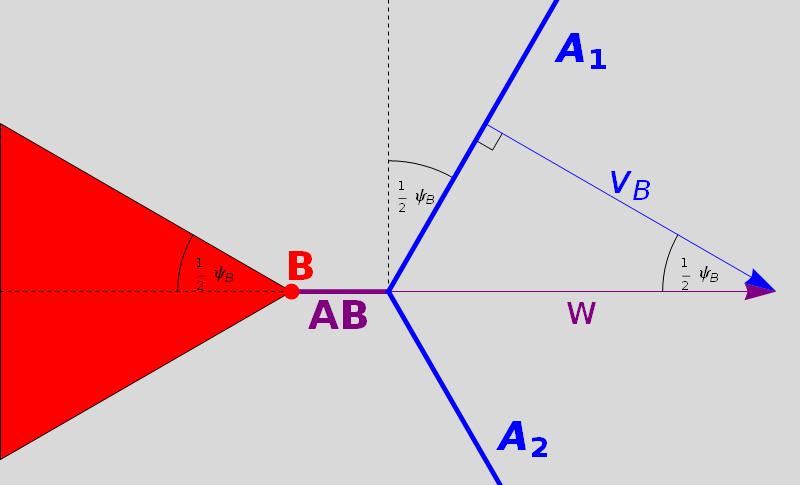}
\caption{The speed $w$ of the junction of the $A_1$, $A_2$, and $AB$ strings measured in the direction of the $AB$ string can be easily obtained using basic geometry.}\label{fig:ABvelocity}
\end{figure}

Since the strings carry some sort of deficit (or surplus) angle, a string that is straight before such a collision cannot be straight after (see figure \ref{fig:simplecollision}). As our model does not allow strings to have kinks --- as such a configuration would imply curvature in its vicinity --- we must add new intermediate strings to complete the configuration.
The simplest situation we can consider is two strings hitting each other at a right angle (see figure \ref{fig:orthogonalcollision}). In that case we can connect the kinks in the two strings with a single finite length string. If we take the paths for the holonomies as shown in the figure, then the holonomy for the intermediate string will satisfy
\begin{equation}\label{eq:intermediatestring}
Q_{AB} =  Q_A Q_B^{-1}Q_A^{-1}Q_B .
\end{equation}
In the rest frame of the $B$ string the velocity of the junction of the $A$ half-strings and the $AB$ finite string measured along the $AB$ string (see figure \ref{fig:ABvelocity})is given by
\begin{equation}
w = \frac{v_A}{\cos(\frac{\psi_B}{2})} ,
\end{equation}
where $v_A$ is the velocity of the $A$ string and $\psi_B$ is the deficit angle of the $B$ string. As $\psi_B$ approaches $\pi$ this velocity will approach infinity. This gives us the first example of the effects of a collision propagating away from the event at superluminal speeds. Now, this superluminal junction is of the least worrisome kind as it represents the two ends of the $B$ strings instantaneously merging to the string $AB$.

In fact, if $w > c$ one can choose a frame in which the two string half-lines $A_1$ and $A_2$ string collide head on. Consequently, one could hope that this seeming non-locality is avoided in more generic collisions. We will therefore proceed to consider more general collisions.

To describe a string we need 7 parameters: the deficit angle $\psi$, two angles to give its orientation, an angle and a positive real number to give the direction and magnitude of its velocity, and two more numbers to give its position with respect to the origin. So, to describe two strings ($A$ and $B$) we need a total of 14 parameters. Of those, 10 can be gauged away by a suitable choice of frame, leaving us with 4 parameters to describe a general collision. A convenient choice is: the deficit angles $\psi_A$ and $\psi_B$ of the two strings, the relative velocities between the strings $v$,\footnote{At times it may be convenient to replace this velocity with the corresponding rapidity $\eta = \arctanh{v/c}$.} and the angle $\phi$ between the two strings at the collision point. 

\begin{figure}[tb]
\centering
\includegraphics[width=120mm]{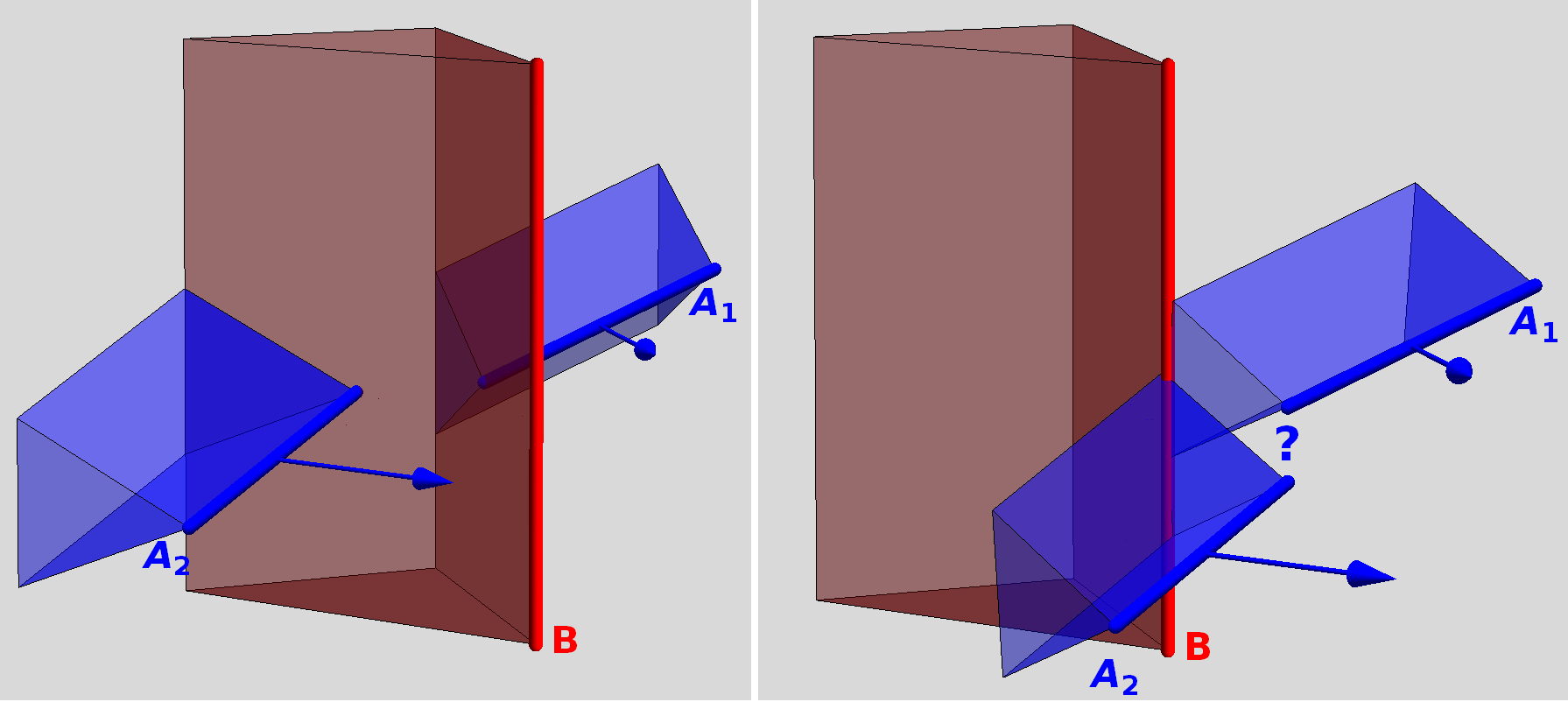}
\caption{When a string $A$ scatters off some other string $B$ at some non-right angle, then the two piece $A_1$ and $A_2$ will not continue to meet at a single point after the collision. It is thus impossible to connect all pieces with just one intermediate string.}\label{fig:nonorthoganalcollision}
\end{figure}

When $\phi$ is not $\pi/2$, it is not possible to resolve the collision with just one intermediate string (see figure \ref{fig:nonorthoganalcollision}). Algebraically, this results from the fact that the holonomy $Q_{AB}$ from equation \ref{eq:intermediatestring}, does not in fact have a frame in which it is a pure rotation, and as such cannot occur as the holonomy of a single string.
\section{Square configurations}\label{sec:squareconfig}
\begin{figure}[btp]
\centering
\includegraphics[width=120mm]{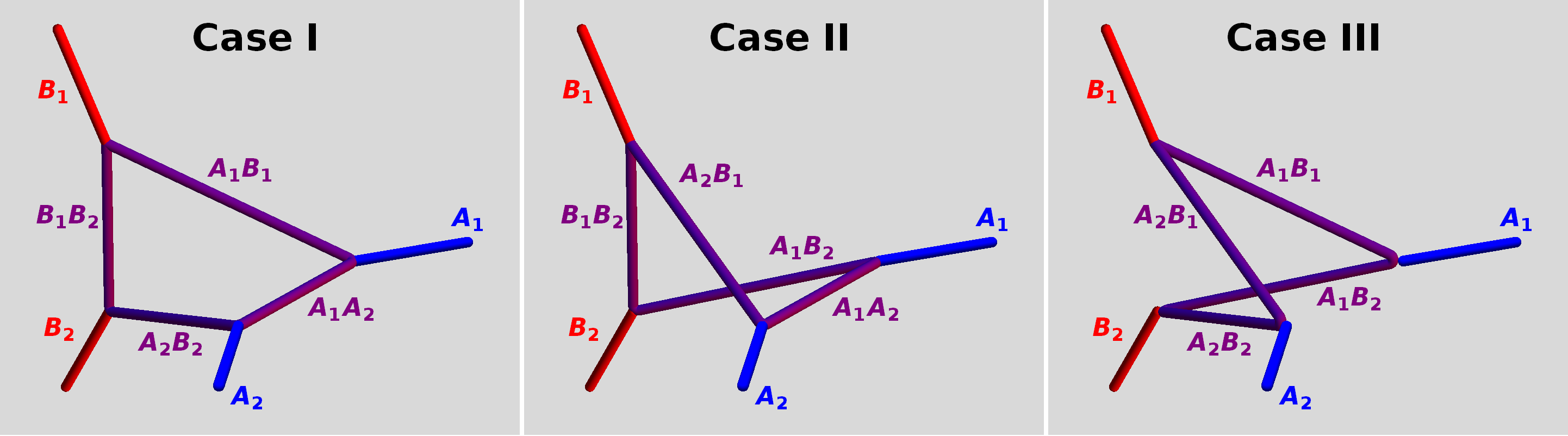}
\caption{There are three different square configurations.}\label{fig:squares}
\end{figure}

In \cite{hooft2008} 't Hooft tried to resolve such a slanted collision by considering a square configuration of four intermediate strings. As shown in figure \ref{fig:squares} there actually are three variants of such a configuration. Here we will focus on case III, because it will turn out to have the simplest solution. The other cases can be solved in a similar way, and have similar (but more complicated solutions).
The four new strings add 28 new parameters to the system. The positions of each of the intermediate strings is fixed by the fact that they must be created in the incidence point of the collision. Consequently, the $(1+1)$-dimensional surface swept out by each of the strings (both internal and external) must pass through this point, which we will take to be the origin of all considered frames allowing us to represent all holonomies as Lorentz transformations. This fixes two parameters of each internal string. 

For suitably chosen paths the conditions to be satisfied by the holonomies at each junction are
\begin{align}\label{eq:squarejunctions}
Q_{A_1B_1} &= Q_{A_1} Q_{A_1B_2} &
Q_{A_2B_1} &= Q_{B_1} Q_{A_1B_1} &
Q_{A_2B_2} &= Q_{A_2} Q_{A_2B_1} &
Q_{A_1B_2} &= Q_{B_2} Q_{A_2B_2} 
\end{align}
with the external holonomies satisfying 
\begin{equation}\label{eq:extcond}
Q_{B_2}Q_{A_2}Q_{B_1}Q_{A_1}=\Id.
\end{equation}
The condition on the external holonomies comes from the fact that they result from the holonomies of two strings before the collision. It also ensures that if any three of the conditions \eqref{eq:squarejunctions} is met, the forth one is automatically also met. The conditions \eqref{eq:squarejunctions} thus represent 18 algebraic conditions on the remaining 20 parameters. It is thus expected that the space of solutions forms a 2 dimensional manifold.

A suitable choice for these parameters are the rapidities $\mu_{A_1}$ and $\mu_{B_1}$ of the junctions on the strings $A_1$ and $B_1$ along those strings. The equations can be solved by representing the Lorentz transformations as elements of $PSL(2,\CC)$.\footnote{We take the Pauli matrices $\sigma_x$, $\sigma_y$, and $\sigma_z$ as generators. A pure rotation of angle $\phi$ along direction $\hat{n}$ is then represented as $\exp(\ii \phi \hat{n}\cdot\vec{\sigma})$, and a pure boost of rapidity $\eta$ in the $\hat{n}$ direction is represented as $\exp( \eta \hat{n}\cdot\vec{\sigma})$.} In this representation, the condition that a holonomy $Q$ represents a string is given by 
\begin{align}\label{eq:tracecond}
-2 <\re \tr &Q < 2;\\
\im \tr &Q =0.
\end{align}

For each junction, we can consider the frame in which the external string is stationary pointing in the $z$-direction. In such a frame the condition on the trace of the internal holonomies implies that each can be written in the form
\begin{equation}
\begin{pmatrix}
a + \ii b & -\ee^{-\mu}(c-\ii d)\\ 
\ee^\mu (c+\ii d)  & a - \ii b
\end{pmatrix},
\end{equation}
with $a^2+b^2+c^2+d^2=1$ and $\mu$ the rapidity of the junction along the external string.\footnote{The value of $\mu$ is ambiguous since we can shift it by an arbitrary boost in the $z$-direction. Here, and elsewhere, we will take $\mu$ to be measured in the frame where the other (colliding) external string has zero velocity in the $z$-direction.} Using this representation we can try to express $\mu_{A_2}$ and $\mu_{B_2}$ in terms of the parameters $\mu_{A_1}$ and $\mu_{B_1}$. The equations typically are a horrendous nonlinear algebraic mess, but by making some clever choice for the involved frames they become manageable for a computer algebra package such as Mathematica.

For case III the result of these calculations is
\begin{align}
\ee^{-2\mu_{A2}}&=
\frac{
\hh{2\frac{\sinh\eta}{\sin\phi}+\tan\frac{\psi_B}{2}\sh{1-\frac{\sinh^2\eta}{{\sin^2\phi}}}}
-\tan\frac{\psi_B}{2}\sh{1+\frac{\sinh^2\eta}{{\sin^2\phi}}}\ee^{-2\mu_{A1}}
}
{
\tan\frac{\psi_B}{2}\sh{1+\frac{\sinh^2\eta}{{\sin^2\phi}}}
+\hh{2\frac{\sinh\eta}{\sin\phi}-\tan\frac{\psi_B}{2}\sh{1-\frac{\sinh^2\eta}{{\sin^2\phi}}}}\ee^{-2\mu_{A1}}
}\label{eq:quadresultA}\\
\ee^{-2\mu_{B2}}&=
\frac{
\hh{2\frac{\sinh\eta}{\sin\phi}+\tan\frac{\psi_A}{2}\sh{1-\frac{\sinh^2\eta}{{\sin^2\phi}}}}
-\tan\frac{\psi_A}{2}\sh{1+\frac{\sinh^2\eta}{{\sin^2\phi}}}\ee^{-2\mu_{B1}}
}
{
\tan\frac{\psi_A}{2}\sh{1+\frac{\sinh^2\eta}{{\sin^2\phi}}}
+\hh{2\frac{\sinh\eta}{\sin\phi}-\tan\frac{\psi_A}{2}\sh{1-\frac{\sinh^2\eta}{{\sin^2\phi}}}}\ee^{-2\mu_{B1}}
}\label{eq:quadresultB}.
\end{align}
Similar results may be obtained for the cases I and II. We notice a couple of things. First of all, the expression for $\mu_{A_2}$ is independent of $\mu_{B_1}$ and the expression for $\mu_{B_2}$ is independent of $\mu_{A_1}$. This odd decoupling of the dependence on the parameters was already noticed in the numerical analysis done in \cite{hooft2008}. This property is common to the solutions of all three cases. The rapidities of two opposite junctions only depend on each other (and the external string parameters).

Second, the relation between $\mu_{A_1}$ and $\mu_{A_2}$ is independent of $\psi_A$ and the relation between $\mu_{B_1}$ and $\mu_{B_2}$ is independent of $\psi_B$. This property is particular to case III and most certainly is related to the fact the opposing junctions lie on the two pieces of the same original string. This is what makes case III easier to deal with since the calculations involve one less parameter. In the other cases the coefficients do depend on both $\psi_A$ and $\psi_B$.

Third, the equations \ref{eq:quadresultA} and \ref{eq:quadresultB} are related to each other by a simple substitution of the labels $A$ and $B$. This simply reflects the original symmetry between the two colliding strings and similar relations hold for the other cases.

Finally (and most crucially), notice that as $\tfrac{\sinh(\eta)}{\sin(\phi)}$ approaches infinity (and $\tan\psi_A$ and $\tan\psi_B$ are positive) the RHS of both equations become negative for any value of the rapidities $\mu_{A_1}$ and $\mu_{B_1}$. The junctions along $A_2$ and $B_2$ thus become superluminal, and unlike the junctions in case of the orthogonal collision we saw before, these are of the most unfavourable kind representing the external strings $A_2$ and $B_2$ instantaneously splitting in two strings. An observer light years away from the collision point could thus instantly become aware of the event. This kind of non-local behaviour cannot be deemed acceptable for any physical theory. One could hope that this behaviour is particular for the considered case (III), and that for any chosen set of values of the collision parameters ($\psi_A$,$\psi_B$,$\phi$, and $\eta$) at least one of the cases I, II, or III would admit a solution with just subluminal junctions. But, alas, the limiting behaviour for the solutions of each of the three cases is the same. For certain values of the collision parameters there are no square string configurations of the resulting state that have only subluminal junctions.

\section{Tetrahedral configurations}\label{sec:tetrahedron}
So there exist values for the collision parameters, for which there are no simple configurations involving one or four intermediate strings that resolve the collision. We can still hope that more involved configurations will provide a suitable resolution.

\begin{figure}\centering
\includegraphics[width=100mm]{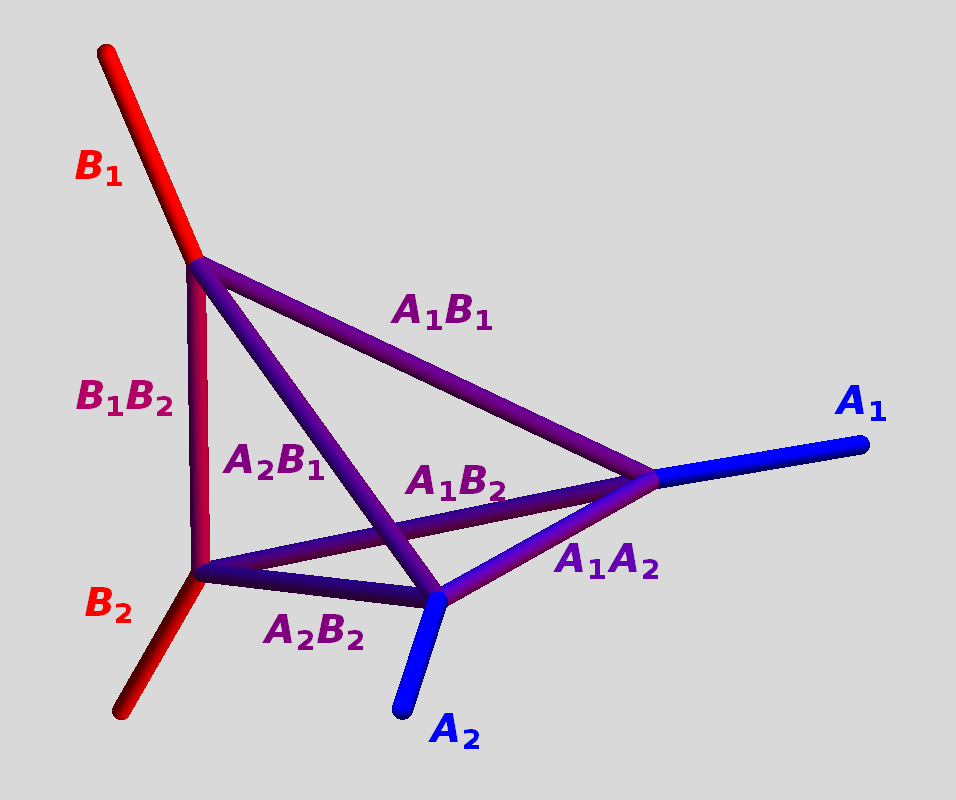}
\caption{Connecting all the vertices yields a tetrahedral configuration of internal strings.}\label{fig:tetrahedron}
\end{figure}

The square configurations considered in the previous section basically consisted of four vertices moving away from the collision  along the external strings which were connected by four internal strings. This involved the choice which vertices should be connected by strings leading to three different cases. We will now consider the more general case where each vertex is connected to each other vertex. We thus end up with a tetrahedral configuration of internal strings (see figure \ref{fig:tetrahedron}). The cases I, II, and III for the square configurations considered before can be considered as special cases of the tetrahedral configuration where the deficit angles of two of the internal strings become zero. For example, case III is obtained by setting the deficit angles of the $A_1A_2$ and $B_1B_2$ strings to zero.

Six internal strings originating from the collision point gives $6\cdot 5=30$ parameters. Of these 18 are fixed by the 18 independent algebraic relations imposed by the algebraic conditions at the vertices, which would leave us with 12 parameters. However, there are extra conditions that result from the fact that we now have 4-vertices. The fact that the algebraic condition on the 4-vertex does not completely fix the 4-vertex, is most easily seen by observing that the algebraic condition for a 4-vertex is the same as the one for a similar configuration of two 3-vertices and an intermediate string (like in figure \ref{fig:orthogonalcollision}). To prevent the 4-vertex splitting up in two 3-vertices we need to require that the velocities of each of the three  internal strings along the external string match. This gives two additional conditions per vertex. The number of free parameters is thus expected to be four.

This agrees with our above observation that we can obtain the square configurations (each having two free parameters) by setting two deficit angles to zero. In case of the square configurations we took two of the velocities of the vertices along the external strings as our free parameters. Since we have four vertices moving along four external strings and four free parameters it is tempting to take these velocities (or their corresponding rapidities) as our free parameters. Among other things this choice allows us to explicitly impose that these vertices move at subluminal speeds.

\subsection{Non-relativistic low energy limit}
Trying to solve the various conditions for the internal parameters of the tetrahedral configuration is very complex. It is thus instructive to first solve these conditions in the limit that all velocities are much smaller than the speed of light (non-relativistic) and all deficit angles are small (all energies are low).

In the low energy limit the string holonomies, that before could be represented by Lorentz transformations, can now be represented by $4\times 4$ Galilean transformations
\begin{equation}
Q =
\begin{pmatrix}
1 & 0 \\
\vec{v} & R(\vec{\phi})
\end{pmatrix},
\end{equation}
where $\vec{v}\in \RR^3$ is the boost velocity and $R(\vec{\phi})\in SO(3)$ is a rotation of $\abs{\vec\phi}$ degrees around the $\vec\phi$ axis. So the holonomy $Q(\vec\phi,\vec{v})$ of a string with orientation $\vec\phi$ and deficit angle $\abs{\vec\phi}$ moving with velocity $\vec{v}$ is given by
\begin{align}
Q(\vec\phi,\vec{v}) &= 
\begin{pmatrix}
1 & 0 \\
-\vec{v} &\Id
\end{pmatrix}
\begin{pmatrix}
1 & 0 \\
0 & R(\vec{\phi})
\end{pmatrix}
\begin{pmatrix}
1 & 0 \\
\vec{v} & \Id
\end{pmatrix}\\
&= \begin{pmatrix}
1 & 0 \\
(\Id-R(\vec{\phi}))\vec{v} & R(\vec{\phi})
\end{pmatrix}\\
&= \begin{pmatrix}
1 & 0 \\
-\vec{\phi}\times\vec{v} & R(\vec{\phi})
\end{pmatrix} + O(\abs{\vec\phi}^2),
\end{align}
where in the last line we used that in the low energy limit  $R(\vec{\phi})\vec{x} = \Id +\vec\phi\times\vec{x}+ O(\abs{\vec\phi}^2)$.

The vertex conditions thus become of the form
\begin{align}
\Id &= Q_1 \cdots Q_n \\
\begin{pmatrix}
1 & 0 \\
0 & \Id \\
\end{pmatrix}
&=
\begin{pmatrix}
1 & 0 \\
-\vec{\phi}_1\times\vec{v}_1 & R(\vec{\phi}_1)
\end{pmatrix}
\cdots
\begin{pmatrix}
1 & 0 \\
-\vec{\phi}_n\times\vec{v}_n & R(\vec{\phi}_n)
\end{pmatrix}+ O(\norm{\vec\phi}^2)\\
&= \begin{pmatrix}
1 & 0 \\
-\vec{\phi}_1\times\vec{v}_1-\ldots-\vec{\phi}_n\times\vec{v}_n & R(\vec{\phi}_1+\dots+\vec{\phi}_n)
\end{pmatrix}+ O(\norm{\vec\phi}^2).
\end{align}
Consequently, in the low energy limit the vertex conditions become
\begin{align}
 \vec{\phi}_1+\dots+\vec{\phi}_n &= 0,\quad\mathrm{and}\label{eq:forcecond}\\
\vec{\phi}_1\times\vec{v}_1+\ldots+\vec{\phi}_n\times\vec{v}_n &= 0.\label{eq:torquecond}
\end{align}

Coming back to our tetrahedral configuration, fixing the velocity $m_i$ of each vertex along the corresponding external string given by $\vec{\phi}_i$ and $\vec{v}_i$ gives us the velocity $\vec{w}_i$ of each vertex
\begin{equation}
 \vec{w}_i = \vec{v_i} + m_i\frac{\vec{\phi}_i}{\norm{\vec{\phi}_i}}.
\end{equation}
These velocities in turn fix the orientations $\hat{\phi}_{i,j}$ and velocities $\vec{v}_{i,j}$ of the internal strings
\begin{align}
 \hat{\phi}_{i,j} &= \frac{\vec{w}_i-\vec{w}_j}{\norm{\vec{w}_i-\vec{w}_j}}, \quad\mathrm{and}\\
\vec{v}_{i,j} &= \frac{\vec{w}_i+\vec{w}_j}{2}.
\end{align}
The only remaining unknown parameters are the deficit angles $\alpha_{i,j}$ of the internal strings. Equation \ref{eq:forcecond} tells us that the $\alpha_{i,j}$ are simply the coefficients of the vector $-\vec{\phi}_j$ when decomposed in the base $\set{\hat{\phi}_{i,j}|i\in\set{A_1,A_2,B_1,B_2} i\neq j}$. It is straight forward to show that this also solves \eqref{eq:torquecond}.

Since we independently obtain values for $\alpha_{i,j}$ and $\alpha_{j,i}$ we may fear that the answer is over determined. However up till now we have ignored the relation \eqref{eq:extcond} for the external strings. These give two relations among the external string parameters
 \begin{align}
 \vec{\phi}_1+\vec{\phi}_2+\vec{\phi}_3+\vec{\phi}_4&= 0,\quad\mathrm{and}\\
\vec{\phi}_1\times\vec{v}_1+\vec{\phi}_2\times\vec{v}_2+\vec{\phi}_3\times\vec{v}_3+\vec{\phi}_4\times\vec{v}_4 &= 0.
\end{align}
It is a straight forward yet involved exercise in linear algebra to show this conditions guarantee that $\alpha_{i,j}= -\alpha_{j,i}$.

We thus find that, in the non-relativistic low energy limit, fixing the velocities of the vertices along the external strings indeed fixes all the internal parameters of the tetrahedral configuration. Since none of the imposed conditions truly degenerate in this limit it is reasonable to assume that this approach should also fix the internal parameters for configurations further away from this limit, although there is absolutely no guarantee that solutions exist for the entire collision parameter space.

\section{A more geometrical approach}\label{sec:cellcomplexes}
Thus far we have tackled the problem of resolving collisions using the holonomies of the involved strings and algebraically solving their relations. Since these relations are typically non-linear, the problem rapidly grows in complexity. The square configuration of intermediate strings was solvable with the help of computer algebra in a reasonable amount of time, but the tetrahedral problem already becomes so complex that it seems intractable with those methods.

However, there is an alternative more geometrical way of describing the situations considered. This will facilitate a much more straight forward way of finding the resolving configuration.

A configuration of strings in spacetime naturally divides the space in a cell complex with each cell having a flat metric.\footnote{This process can require the adding of extra strings with zero tension/deficit angle, which can also be tachyonic since they do not represent any physical information. There may be multiple ways in which such strings can be added, making the cell structure for a given configuration of strings not unique. If desired the addition of strings can be continued to make the cell complex into a simplicial complex.} Collision points form the $0$-cells of this complex, junctions the $1$-cells, and the strings themselves form the $2$-cells.

Multiple strings can form loops like the square in the configuration considered in section \ref{sec:squareconfig}.  Typically such a loop does not lie in a single hyperplane, but we can add new strings to divide the loop in to smaller loops that do. These `flat' loops will become the $3$-cells of the complex. The new virtual strings\footnote{From here on we will use the term `virtual string' to refer to strings that carry trivial holonomy.} added this way have trivial holonomy, and thus are not subject to the physical limitation of being timelike. Typically, there are multiple ways in which virtual strings may be added to obtain flat $3$-cells. The cell structure obtained is thus not unique.

The system of $3$-cells will divide $(3+1)$-dimensional space in disjoint pieces. Each such piece will become a $4$-cell in the cell complex.

Each $n$-cell in the complex will inherit a flat metric from the original spacetime. Moreover, the attaching maps are simply the inclusion of the boundary of each $n$-cell in the skeleton of $(n-1)$-cells. In particular, the attaching maps will preserve the flat metrics of the cells, and are thus given (piecewise) as Poincar\'{e} transformations.

Conversely, given such a cell complex with flat metrics we can try to reconstruct a string configuration. This requires some conditions on the cell complex. First of all the cell-complex must be a orientable topological 4-manifold.\footnote{This involves some technical conditions on the cell complex. Each $3$-cell should be attached to exactly two $4$-cells, each $(n-1)$-cell is connected to at least 2 $n$-cells. The fourth homology group over the real numbers should be isomorphic to $\RR$. Et cetera.} Moreover, the flat metric of each $4$-cell must have Lorentzian signature.

Since each $4$-cell has a flat Lorentzian metric, we can assign a global Poincar\'{e} frame to each $4$-cell. Then if $4$-cells $\alpha$ and $\beta$ are attached to a single $3$-cell the attachment maps will induce a Poincar\'{e} transformation $M_{\alpha\beta}$ mapping the frame of $\alpha$ to the frame of $\beta$. The holonomy of a $2$-cell can then be determined by composing the  $M_{\alpha\beta}$'s of all the $3$-cells attached to it.

If a $2$-cell has a non-trivial holonomy it must represent a string. In order to avoid unphysical tachyonic strings, we must require that all $2$-cells with non-trivial holonomy, have a metric with Lorentzian signature (i.e. propagate at subluminal speeds.) Finally, in order to avoid superluminal junctions we can explicitly require that the inclusion of each $1$-cell that is attached to 3 or more physical strings (i.e. $2$-cells with non-trivial holonomy) into a physical string is timelike.

So, we are basically describing Regge manifolds with some extra conditions connected to the requirement that the strings behave as physical excitations.

\subsection{Back to tetrahedral configurations}\label{sec:cellresolution}
\begin{figure}
\centering
\includegraphics[width=100mm]{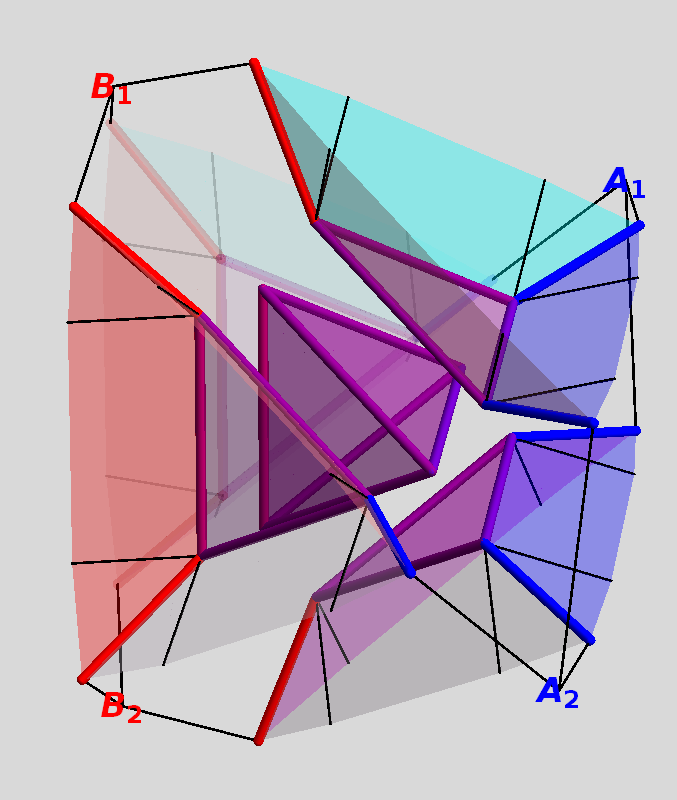}
\caption{The tetrahedral configuration can be divided in five $4$-cells. Here shown on a time slice after the collision, such that each $4$-cell is represented by a $3$-dimensional cell. There are four external cells extending to infinity, and one internal tetrahedral cell. Surfaces (representing $3$-cells) with matching colour are mapped into each other through a Lorentz transformation. The black lines on the external surfaces are virtual strings that subdivide the external surfaces and carry no holonomy. These are necessary because the external strings generically do not lie in the same plane.} \label{fig:tetracells}
\end{figure}

With this more geometrical approach, we can try to tackle the tetrahedral resolution of the string collision.  We will divide the tetrahedral configuration up in five cells as shown in figure \ref{fig:tetracells}. 

For this we start out far away from (outside the light cone of) the collision. Here the geometry of the spacetime is not yet effected by the collision and can be constructed by continuing the geometry as it existed before the collision. In this region there will be four half-strings pointing in towards the collision, each a half of one of the two original strings.

This region can be roughly divided into four 4-cells, each corresponding to an area that is in between three of the half-strings and opposite to the fourth. Since two half-strings will typically not lie in a single plane, it is necessary to add virtual strings to properly define the boundary between the `external' 4-cells as flat 3-cells.

In each of these external 4-cells we can continue the geometry inward towards the collision until we reach the junctions on each of the half-strings. A priori, we are still free to set the speeds of these junctions.\footnote{We will return to the question what velocities can be chosen in the next section.} Particularly, we take these speeds to be subluminal. Since all the junctions originate in the point of collision, the three junctions (1-cells) in an external 4-cell will lie in a single 3-plane. This 3-plane will become the inner boundary 3-cell of the external 4-cell. The boundary of this 3-cell will consist of three 2-cells each spanned by two of the junctions. Since each of the junctions is subluminal, these 2-cells are timelike, and we can thus interpret them as the internal strings of the tetrahedral configuration.

We thus know the geometry on the external $4$-cells and the entire $3$-skeleton of the tetrahedral configuration. The missing piece to be added is a 4-cell with a piecewise flat metric, which is to be attached to the boundary formed by the four interior boundary 3-cells of the external 4-cells.  The geometry of each of the boundary 3-cells is simply that of a triangle that grows linearly with time. The geometry of the internal 4-cell should thus be that of a tetrahedron filled with a piecewise flat metric that expands linearly with time and that agrees with the given flat metrics on the boundary triangles.

The problem of filling a tetrahedron with an internal flat metric given a flat metric on the boundary is well-known in Euclidean geometry. It is known that this is possible if and only if the boundary metrics satisfy the (generalized) triangle inequalities.\cite{Schoenberg1935} That is, if and only if the sum of the areas of any three of the boundary triangles is larger than the area of the remaining triangle.

It is not clear to us that these triangle inequalities will be satisfied for any choice of the collision parameters and junction speeds. We therefore assume the worst possibility, that there exist situations in which these are not satisfied. Consequently, we must provide a way to construct an internal piecewise flat metric. This is indeed possible for any flat geometry of the boundary.

\begin{figure}
\centering
\includegraphics[width=70mm]{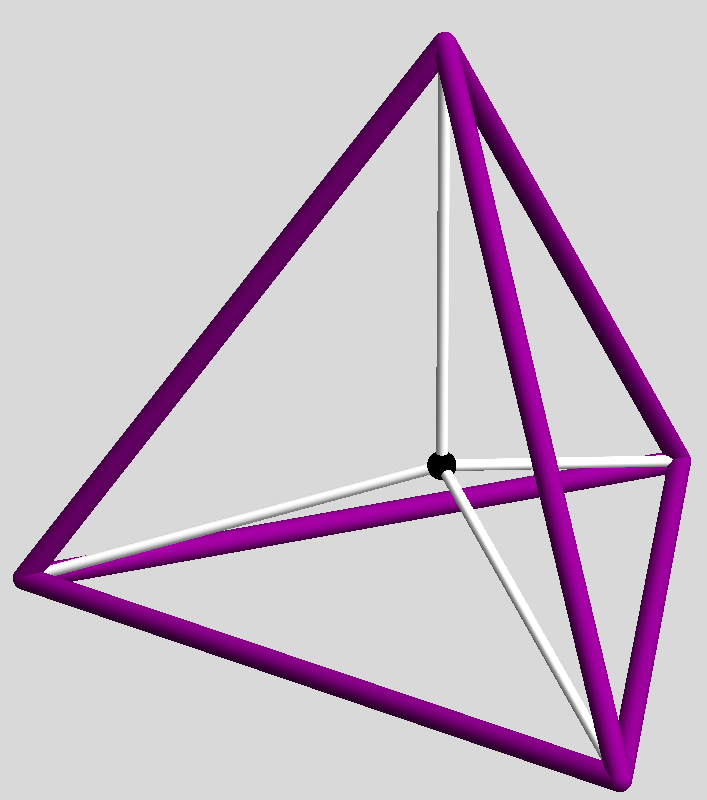}
\caption{A tetrahedron can be subdivided into four smaller tetrahedra by adding a single vertex, four 1-cells, six 2-cells, and four 3-cells.}\label{fig:subdivide}
\end{figure}

To construct an internal piecewise flat metric in the tetrahedron, we will subdivide it in four pieces (see figure \ref{fig:subdivide}). We build a cell complex (in fact a simplicial complex) on the interior in the following way. We add a single vertex, which we connect to each of the four vertices on the boundary by 1-cells. Each of the triangles formed by two of the new internal 1-cells and one of the 1-cells on the boundary is filled by a new 2-cell, and each of the tertahedra formed by three of the new 2-cells and one of the boundary 2-cells is filled by a new 3-cell.

For each of the new cells we will have to specify a flat metric that is compatible with the metrics on its boundary. Any metric on a 1-cell is flat and is specified by a single parameter; its length. Consequently, the four new internal 1-cells give us four free parameters. To construct a flat metric on a 2-cell we need that the metrics on its boundary 1-cells satisfy the (2D) triangle inequalities. This can easily be satisfied for all the new 2-cells. One particular way is to choose all four lengths equal and larger than half the length of the longest 1-cell on the boundary.
 
To construct a flat metric on a 3-cell, we again need the metrics on its boundary 2-cells to satisfy the (3D) triangle inequalities. The choice above guarantees that three of the four inequalities are satisfied. 

To see this, call the length of the new internal 1-cells $a$ and the length of the three 1-cells on the outer boundary of any particular internal 3-cell $b_1$, $b_2$, and $b_3$. Because they are on the boundary of a given triangle they satisfy
\begin{equation}
 b_i + b_j \geq b_k,
\end{equation}
with $i,j,k\in \set{1,2,3}$, and without loss of generality we can take the labelling such that $b_1 \leq b_2 \leq b_3$. The area $A_i$ of the internal 2-cell incident to the 1-cell with length $b_i$ is thus equal to
\begin{equation}
A_i = \frac{1}{2}b_i\sqrt{a^2- \frac{1}{4}b_i^2}.
\end{equation}
Since $a> \frac{1}{2}b_i$ for all $i$, $b_1 \leq b_2 \leq b_3$ implies that $A_1 \leq A_2 \leq A_3$. Furthermore we have
\begin{align}
 A_1 + A_2 &=  \frac{1}{2}b_1\sqrt{a^2- \frac{1}{4}b_1^2} + \frac{1}{2}b_2\sqrt{a^2- \frac{1}{4}b_2^2} \\
  &\geq \frac{1}{2}(b_1+b_2)\sqrt{a^2- \frac{1}{4}b_3^2}\\
  &\geq \frac{1}{2}(b_3)\sqrt{a^2- \frac{1}{4}b_3^2} 
  &\geq A_3.
\end{align}
As a result any sum involving two of the areas of the internal 2-cells will be larger than the area of the remaining 2-cell.

The remaining triangle inequality is that the sum of the areas of the internal 2-cells is larger  than the area of the boundary 2-cell. Since the areas of the internal 2-cells can be made arbitrarily large by increasing $a$, it is possible to also satisfy this fourth triangle inequality for the new 3-cells.

Hence we can construct a piecewise flat metric on the interior of the tetrahedron. Consequently, the geometry of the missing 4-cell  can be taken to be such a piecewise flat metric expanding linearly with time. That is, we can expand our original cell complex by adding four new internal 4-cells each with the geometry of a tetrahedron growing linearly with time (each corresponding to one of the internal 3-cells in our construction of the piecewise flat metric).

The new 1-cells in the construction of the piecewise flat metric, correspond to four new internal 2-cells that are timelike and generically carry nontrivial holonomy. These must therefore be interpreted as new internal strings.

As is clear from the arbitrary choices we made along the way this construction is far from unique.
 
\subsection{Limits}\label{sec:limits}
So does this construction always work? In the exposition above we have sidestep\-ped a couple of possible hurdles. The first issue is that due to the presence of the holonomies of the colliding strings the geometry around the collision point is non-Euclidean. In particular two lines passing through a single point  are not guaranteed to have a single plane connecting the two. There maybe more planes connecting the two (which is not really a problem but does add another arbitrary choice) or worse there may be none at all. This is bad since it  prevents us from connecting the junctions with strings.

\begin{figure}
\centering
\includegraphics[width=120mm]{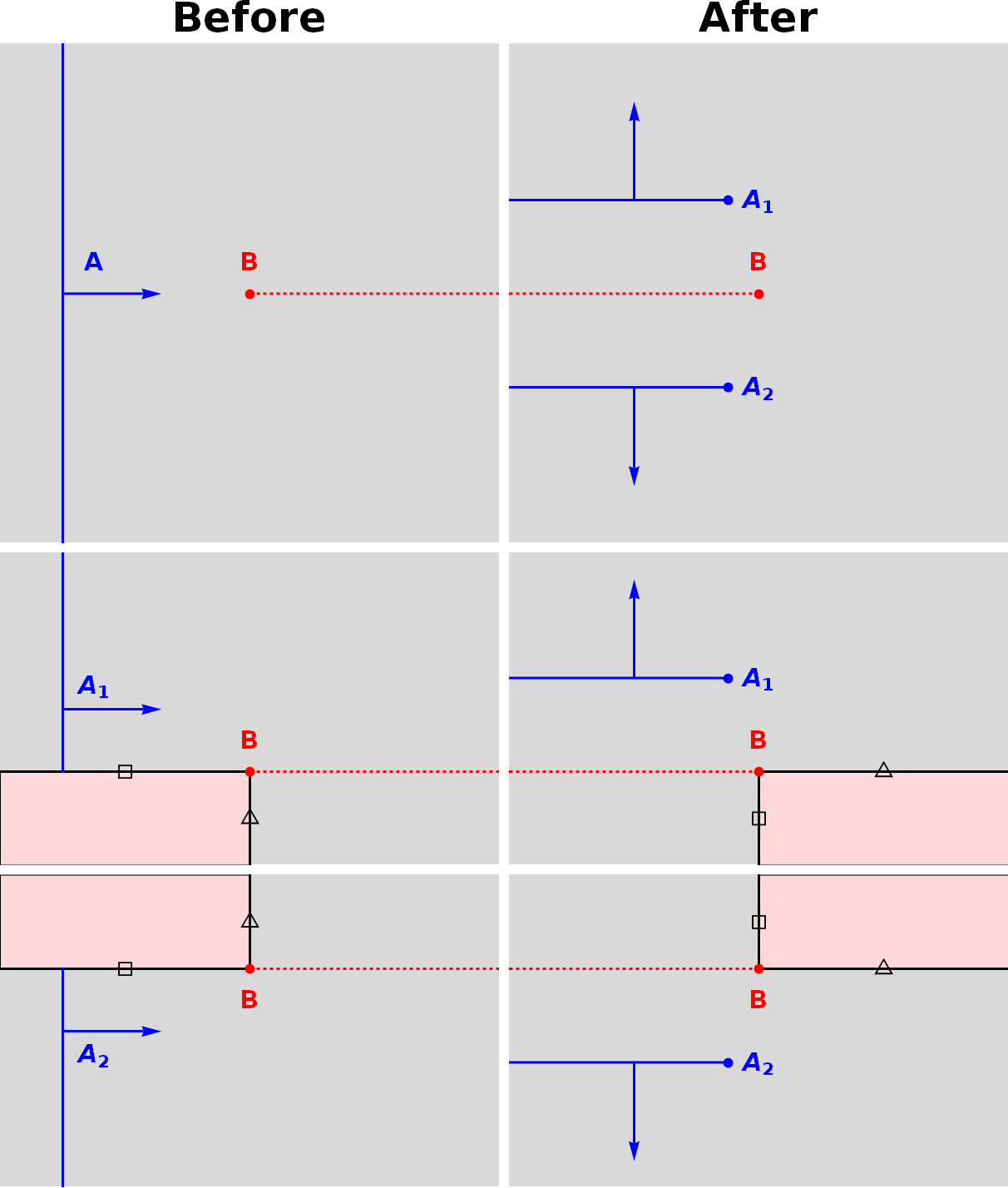}
\caption{String $A$ is scattering off the stationary string $B$, which has a surplus angle of $\pi$. (Viewed along the direction of $B$.) The dashed line indicates the cut leading to the surplus area. On the bottom row the same process is depicted, but now as two separate pieces of space that are glued together along boundary of the red area. (The lines marked with a triangle are identified with each other, idem for the lines marked with a square.) After the collision there exist no straight lines going from $A_1$ to $A_2$. }\label{fig:largesurplus}
\end{figure}

The latter can indeed happen if one of the original strings has a very large surplus angle. For example, consider the case in figure \ref{fig:largesurplus}, where a string is colliding with a second stationary string with a surplus angle of $\pi$. After the collision, no matter what frame you choose, there is no way to draw a straight line between the points $A_1$ and $A_2$. You might try to solve this by allowing the endpoints of $A_1$ and $A_2$ to move in the negative direction, but  the issue will just reappear at even larger surplus angles and larger incident speeds.
Large surplus angles can thus pose serious issues for finding our type of solution. This suggests that we need to avoid surplus angles, or at least control them in such a way that they do not become arbitrarily large. 

\begin{figure}
\centering\includegraphics[width=80mm]{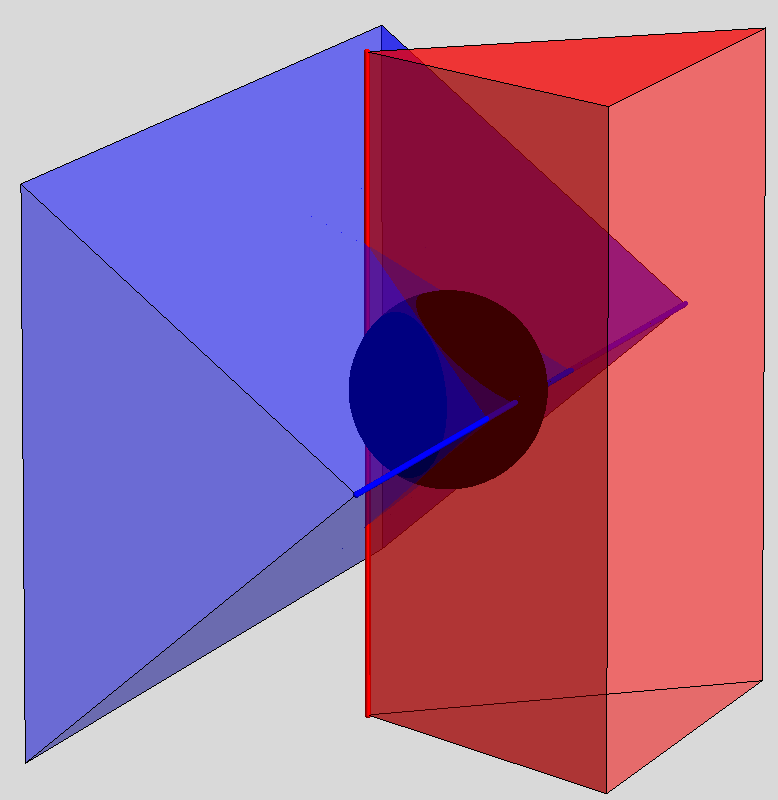}
\caption{Two strings colliding at high velocity viewed from the centre of velocity frame. The black sphere in the middle marks all the points moving away from the collision point at the speed of light. The area is completely cut out by the deficiency angles of the two strings.}\label{fig:cutsphere}
\end{figure}

Another convenient assumption we made was that the junctions can always be chosen to move at a subluminal speed. We already know that this is a dangerous assumption to make since this was exactly the thing that failed in previous attempts to identify resolutions. And in fact it also fails here. In the construction in section \ref{sec:cellresolution} it is relatively easy to find an example where junctions can never be subluminal. This generally happens when both strings have large deficiency angles. Figure \ref{fig:cutsphere} shows the result of continuing the geometry of two strings colliding almost orthogonally at high velocity, as viewed from the centre of velocity frame\footnote{i.e. the frame where the sum of the velocities of the strings vanishes.}.
The black sphere marks all the points moving away from the collision at the speed of light. The sphere is completely contained in the wedges indicating the area removed by the deficiency angles of the two strings. Since any junction on the strings will have to be outside the wedges it must move at a superluminal speed.

This example not only shows that it is impossible to resolve a collision with a tetrahedral configuration of intermediate strings, but it squashes all hope that any more complicated configuration of internal strings can do the job while all junctions stay subluminal.

From the same centre of velocity frame we can obtain a safe limit on the collision parameters for which the resolution using a tetrahedral configuration of strings will always work. If the half-string of string $B$ meets the sphere of points moving away from the collision point at the speed of light before it meets the wedge cut out by the $A$ string, then we can find a subluminal location for the junction on that half-string. If we take the deficiency angles of the strings (in their respective rest-frames) to be $\psi_A$ and $\psi_B$, the angle between the strings at the collision $\phi$, and the speed of the string with respect to the collision point $v$, then this condition can be expressed as
\begin{equation}
(2v)^2 +\hh{2v\gamma(v)\frac{\tan\psi_A/2}{\sin\phi}}^2 \leq 1,
\end{equation}
where the Lorentz factor $\gamma(v)$ appears due to the Lorentz contraction of the moving wedge of string $A$. If we impose this condition for both strings and in addition require both deficiency angles to be positive, we can chose all four junctions to move at the speed of light and connect them by intermediate strings to find the geometry of the boundary of the tetrahedron. Moreover due to the symmetry of the situation, the pair of opposing faces incident to each string will have the same area. This guarantees that the triangle inequalities are satisfied and we can thus find a flat interior for the tetrahedron. We thus obtain as a safe condition on the collision parameters that each $\psi$ must satisfy
\begin{equation}
0< \tan\frac{\psi}{2} \leq \sqrt{\sin^2\phi\frac{1-4v^2}{4v^2 \gamma(v)}}.
\end{equation}
Within those bounds it is clear that the construction will always work. This bound may not be maximal in the sense that there may be values beyond these bounds for which a resolution may still be obtained for either the tetrahedral of a more complicated configuration of internal strings. But, this is not guaranteed and it is clear from the examples mentioned above that at some point beyond these bounds it will become impossible to find a resolution.

\section{Comparison to other piecewise flat gravity approaches}\label{sec:comparison}
At the first sight the model proposed by 't Hooft is similar to many other approaches describing gravity that is piecewise flat. We will here discuss these similarities and point out some notable differences.

As a first remark note that the dynamical model studied here is not quantum mechanical, so it must be considered as the classical limit of a discretized theory. Our study of it is motivated by the wish to understand what the rules are for both a classical and a quantized discrete Regge-like model, its space of states, the question of the positivity of the energy, et cetera. Other models, on the other hand, often delve directly into the quantum aspects of such a model. 

The proposed  model has a lot in common with the world crystal model proposed by Kleinert. \cite{Kleinert:2003za,Kleinert2008} That model also proposes to describe gravity through propagating topological defect lines. The difference is that his defect lines are not necessarily straight nor follow a constant trajectory. This means that he is not imposing that Einstein's equation should hold on top of the requirement that empty space is (locally) flat. It is precisely the combination of these two requirements that implies that strings must have planar world sheets.

It also seems that there might be some connection with loop quantum gravity approaches and the related spin foam models. These also take holonomies as fundamental variables, and have historically arisen from considerations of Ponzano-Regge models, which also feature piecewise flat manifolds.  One could wonder for example if the model considered here appears as the classical limit of LQG. This question was addressed recently by Eugenio Bianchi\cite{bianchi2009}, who came to the conclusion that to reproduce the kinematical state space of LQG by standard path integral quantization techniques, one should consider the holonomies of loops around string defects caused by locally-flat connections. This is a weaker condition than restriction to locally-flat metrics made by us here.

Another similarity is with causal dynamical triangulations. \cite{Ambjorn:1998xu,Ambjorn:2000dv,Ambjorn:2001cv} The configurations of 4-simplices with a Lorentzian metric pasted together considered there are very similar to the cell complexes described in section \ref{sec:cellcomplexes}. A cell complex can always be reduced to a simplicial complex by repeatedly dividing the cells until all cells are simplices. The difference here is the emphasis we lay on considering the 2-cells with nontrivial holonomy as physical degrees of freedom resulting in the requirement that all such cells be timelike. CDT makes no such requirement on its configurations. By construction it contains many spacelike 2-simplices that generically do not have a trivial holonomy.

\section{Conclusions}
We have revisited the locally flat gravity model introduced by 't Hooft  and studied the problem of resolving the collisions of flat strings. We have found a closed form solution for the quadrangle resolutions proposed in his paper, which confirms his conclusion based on numerical analysis that this resolution becomes incompatible with the requirements of causality for certain values of the collision parameters. 

In the hope of finding resolutions for these situations we have introduced a new more complicated configuration with six internal strings forming a tetrahedron. We have shown that in the non-relativistic low energy limit this resolution is solvable for any choice of the collision parameters with the speeds of the junctions of the tetrahedron along the external strings as free parameters. To further analyse this configuration we have introduced the description if the configuration as a piecewise flat manifold as an alternative the algebraic description using the Poincar\'e holonomies.

Using this new geometric description we have shown how to find a tetrahedral resolution of a collision. To guarantee that the junction are subluminal and thus do not violate causality we need to restrict to a bounded range of the collision parameters. Beyond this range it is not clear that resolutions exists that satisfy causality, and in fact there are examples of the collision parameters for which it is clear that no resolution---no matter how complicated---exists that satisfies this condition. A viable model with straight strings must thus only contain collisions that satisfy certain bounds on the collision parameters. Whether it is possible for such a model to be consistent is a possible future line of enquiry. 

\section*{Acknowledgements}
The author would like his advisor, Gerard 't Hooft, for his continuing encouragements and his corrections and suggestions.

\bibliographystyle{utcaps}
\bibliography{strings}

\providecommand{\href}[2]{#2}\begingroup\raggedright\begin{thebibliography}{10}

\bibitem{hooft2008}
G.~'t~Hooft, ``{A locally finite model for gravity},''
  \href{http://dx.doi.org/10.1007/s10701-008-9231-3}{{\em Found. Phys.} {\bf
  38} (2008)  733--757}, \href{http://arxiv.org/abs/0804.0328}{{\tt
  arXiv:0804.0328 [gr-qc]}}.

\bibitem{Deser:1983tn}
S.~Deser, R.~Jackiw, and G.~'t~Hooft, ``{Three-Dimensional Einstein Gravity:
  Dynamics of Flat Space},''
  \href{http://dx.doi.org/10.1016/0003-4916(84)90085-X}{{\em Ann. Phys.} {\bf
  152} (1984)  220}.

\bibitem{'tHooft:1993gz}
G.~'t~Hooft, ``{The Evolution of gravitating point particles in (2+1)-
  dimensions},''
{\em Class. Quant. Grav.} {\bf 10} (1993)  1023--1038.

\bibitem{'tHooft:1996uc}
G.~'t~Hooft, ``{Quantization of Point Particles in 2+1 Dimensional Gravity and
  Space-Time Discreteness},''
  \href{http://dx.doi.org/10.1088/0264-9381/13/5/018}{{\em Class. Quant. Grav.}
  {\bf 13} (1996)  1023--1040},
\href{http://arxiv.org/abs/gr-qc/9601014}{{\tt arXiv:gr-qc/9601014}}.

\bibitem{Kadar:2004im}
Z.~Kadar, ``{Polygon model from first order gravity},''
  \href{http://dx.doi.org/10.1088/0264-9381/22/5/004}{{\em Class. Quant. Grav.}
  {\bf 22} (2005)  809--824},
\href{http://arxiv.org/abs/gr-qc/0410012}{{\tt arXiv:gr-qc/0410012}}.

\bibitem{Witten:1988hc}
E.~Witten, ``{(2+1)-Dimensional Gravity as an Exactly Soluble System},''
\href{http://dx.doi.org/10.1016/0550-3213(88)90143-5}{{\em Nucl. Phys.} {\bf
  B311} (1988)  46}.

\bibitem{Schoenberg1935}
I.~J. Schoenberg, ``Remarks to Maurice Frechet's Article ``Sur La Definition
  Axiomatique D'Une Classe D'Espace Distances Vectoriellement Applicable Sur
  L'Espace De Hilbert,'' {\em Ann. Math.} {\bf 36} (1935) no.~3, 724--732.
  \url{http://www.jstor.org/stable/1968654}.

\bibitem{Kleinert:2003za}
H.~Kleinert and J.~Zaanen, ``{World nematic crystal model of gravity explaining
  the absence of torsion},''
  \href{http://dx.doi.org/10.1016/j.physleta.2004.03.048}{{\em Phys. Lett.}
  {\bf A324} (2004)  361--365},
\href{http://arxiv.org/abs/gr-qc/0307033}{{\tt arXiv:gr-qc/0307033}}.

\bibitem{Kleinert2008}
H.~Kleinert, {\em Multivalued Fields}.
\newblock World Scientific, 2008.

\bibitem{bianchi2009}
E.~Bianchi, ``{Loop Quantum Gravity a la Aharonov-Bohm},''
  \href{http://arxiv.org/abs/0907.4388}{{\tt arXiv:0907.4388 [gr-qc]}}.

\bibitem{Ambjorn:1998xu}
J.~Ambjorn and R.~Loll, ``{Non-perturbative Lorentzian quantum gravity,
  causality and topology change},''
  \href{http://dx.doi.org/10.1016/S0550-3213(98)00692-0}{{\em Nucl. Phys.} {\bf
  B536} (1998)  407--434},
\href{http://arxiv.org/abs/hep-th/9805108}{{\tt arXiv:hep-th/9805108}}.

\bibitem{Ambjorn:2000dv}
J.~Ambjorn, J.~Jurkiewicz, and R.~Loll, ``{A non-perturbative Lorentzian path
  integral for gravity},''
  \href{http://dx.doi.org/10.1103/PhysRevLett.85.924}{{\em Phys. Rev. Lett.}
  {\bf 85} (2000)  924--927},
\href{http://arxiv.org/abs/hep-th/0002050}{{\tt arXiv:hep-th/0002050}}.

\bibitem{Ambjorn:2001cv}
J.~Ambjorn, J.~Jurkiewicz, and R.~Loll, ``{Dynamically triangulating Lorentzian
  quantum gravity},''
  \href{http://dx.doi.org/10.1016/S0550-3213(01)00297-8}{{\em Nucl. Phys.} {\bf
  B610} (2001)  347--382},
\href{http://arxiv.org/abs/hep-th/0105267}{{\tt arXiv:hep-th/0105267}}.

\end{thebibliography}\endgroup
\end{document}